\documentclass[twoside]{article}
\usepackage{amsfonts,amssymb,amsbsy,textcomp,marvosym,picins,amsmath,caption,threeparttable,amsthm,subfigure,float,lastpage,lscape}
\usepackage{eurosym,mathrsfs,fancyhdr,CJK,multicol,graphics,indentfirst,color,bm,upgreek,booktabs,graphicx,multirow,warpcol,epstopdf,algorithm,algorithmic}
\usepackage[bookmarksnumbered=true,bookmarksopen=true,colorlinks=true,pdfborder=001,urlcolor=blue,linkcolor=blue,anchorcolor=blue,citecolor=blue]{hyperref}
\usepackage{orcidlink}

\usepackage[numbers,sort&compress,super]{natbib} 
\setcitestyle{open={[},close={]}}
\renewcommand{\thefootnote}{\fnsymbol{footnote}}
\def\footnoterule{\kern 1mm \hrule width 10cm \kern 2mm}

\newcommand{\secref}[1]{\hyperref[#1]{Section \ref*{#1}}}
\newcommand{\subsecref}[1]{\hyperref[#1]{Subsection \ref*{#1}}}
\newcommand{\figref}[1]{\hyperref[#1]{Fig.\ref*{#1}}}
\newcommand{\tableref}[1]{\hyperref[#1]{Table \ref*{#1}}}
\newcommand{\equref}[1]{\hyperref[#1]{(\ref*{#1})}}
\allowdisplaybreaks
\catcode`@=11
\def\title#1{\vspace{3mm}\begin{flushleft}\vglue-.1cm\Large\bf\boldmath\protect\baselineskip=18pt plus.2pt minus.1pt #1
\end{flushleft}\vspace{1mm} }
\def\author#1{\begin{flushleft}\normalsize #1\end{flushleft}\vspace*{-4pt} \vspace{3mm}}
\def\address#1#2{\begin{flushleft}\vglue-.35cm${}^{#1}$\small\it #2\vglue-.35cm\end{flushleft}\vspace{-2mm}\par}
\def\jz#1#2{{$^{\footnotesize\textcircled{\tiny #1}}$\footnotetext{$^{\footnotesize\textcircled{\tiny #1}}$#2}}}

\catcode`@=11
\def\section{\@startsection{section}{1}{\z@}%
 {-3ex \@plus -.3ex \@minus -.2ex}%
 {2.2ex \@plus.2ex}%
{\normalfont\normalsize\protect\baselineskip=14.5pt plus.2pt minus.2pt\bfseries}}
\def\subsection{\@startsection{subsection}{2}{\z@}%
 {-3ex\@plus -.2ex \@minus -.2ex}%
 {2ex \@plus.2ex}%
{\normalfont\normalsize\protect\baselineskip=12.5pt plus.2pt minus.2pt\bfseries}}
\def\subsubsection{\@startsection{subsubsection}{3}{\z@}%
 {-2.2ex\@plus -.21ex \@minus -.2ex}%
 {1.4ex \@plus.2ex}
{\normalfont\normalsize\protect\baselineskip=12pt plus.2pt minus.2pt\sl}}

\begin{document}
\begin{CJK*}{GBK}{song}
\thispagestyle{empty}
\vspace*{-13mm}
\noindent {\small Liu PS, Zheng LN, Chen JL {\it et al.} Enhancing Recommendation with Denoising Auxiliary Task. JOURNAL OF COMPUTER SCIENCE AND TECHNOLOGY \ 33(1): \thepage--\pageref{last-page}\ Jun. 2024. DOI: 10.1007/s11390-024-4069-5}
\vspace*{2mm}

\title{Enhancing Recommendation with Denoising Auxiliary Task}

\author{Pengsheng Liu\orcidlink{0009-0002-4406-7410}$^{1,\spadesuit}$ , Linan Zheng\orcidlink{0009-0004-4437-1556} $^{2,3,\spadesuit}$ , Jiale Chen$^{2,3}$ , Guangfa Zhang$^{2,3}$ , Yang Xu$^{1}$ , Jinyun Fang$^{2,*}$ }

\address{1}{College of Big Data and information engineering, Guizhou University, Guiyang 550025, China}
\address{2}{Institute of Computing Technology, Chinese Academy of Sciences, Beijing 100190, China}
\address{3}{University of the Chinese Academy of Sciences, Beijing 100049, China}

\vspace{2mm}

\noindent E-mail: ie.psliu21@gzu.edu.cn; zhenglinan21@mails.ucas.ac.cn; chenjiale21s@ict.ac.cn; zhangguangfa14@mails.ucas.ac.cn; xuy@gzu.edu.cn; fangjy@ict.ac.cn;\\[1mm]

\noindent Received December 25, 2023; accepted June 25, 2024.\\[1mm]

\let\thefootnote\relax\footnotetext{{}\\[-4mm]\indent\ This is a post-peer-review, pre-copyedit version of an article published in Journal of Computer Science and Technology at \url{https://jcst.ict.ac.cn/en/article/doi/10.1007/s11390-024-4069-5}\\[.5mm]
\indent\ The work was supported by the Program for Student Innovation through Research and Training under Grant No. 2023SRT071.\\[.5mm]
\indent\ $^\spadesuit$Equal Contributions\\[.5mm]
\indent\ $^*$Corresponding Author\\[.5mm]
\indent\ \copyright Institute of Computing Technology, Chinese Academy of Sciences 2024}

\noindent {\small\bf Abstract} \quad  {\small {The historical interaction sequences of users plays a crucial role in training recommender systems that can accurately predict user preferences. However, due to the arbitrariness of user behavior, the presence of noise in these sequences poses a challenge to predicting their next actions in recommender systems. To address this issue, our motivation is based on the observation that training noisy sequences and clean sequences (sequences without noise) with equal weights can impact the performance of the model. We propose a novel self-supervised Auxiliary Task Joint Training (ATJT) method aimed at more accurately reweighting noisy sequences in recommender systems. Specifically, we strategically select subsets from users' original sequences and perform random replacements to generate artificially replaced noisy sequences. Subsequently, we perform joint training on these artificially replaced noisy sequences and the original sequences. Through effective reweighting, we incorporate the training results of the noise recognition model into the recommender model. We evaluate our method on three datasets using a consistent base model. Experimental results demonstrate the effectiveness of introducing self-supervised auxiliary task to enhance the base model's performance.}}

\vspace*{3mm}


\vspace*{4mm}

\end{CJK*}
\baselineskip=14.8pt plus.2pt minus.2pt
\parskip=0pt plus.2pt minus0.2pt
\begin{multicols}{2}

\section{Introduction}
Recommender systems play a crucial role in today's internet and e-commerce domains, offering users improved information retrieval and shopping experiences, while also yielding substantial economic benefits for businesses \cite{1,2,3}. Click-through rate (CTR) prediction holds a significant role within personalized recommender systems \cite{4,5,6,7}. By analyzing users' historical interaction sequences, these systems recommend products aligned with user interest and preferences, facilitating the discovery of potentially engaging content \cite{8}. This method enhances user experience, fosters sales and propagates content \cite{9,10}.
In the context of sequence-based recommendation, the issue of noise present in sequences significantly impacts the establishment of accurate and reliable recommender models, forming a complex and pivotal challenge within the field. Sequence noise can arise from various sources, including user curiosity, data collection inaccuracies and environmental shifts, consequently leading to misjudgments of user interest and inaccurate recommender model outcomes \cite{11,12,13,14,15,16,17,7}. Models trained on clean sequences significantly outperform those trained on original, noise-containing sequences. This underscores the imperative of exploring denoising strategies in recommender systems \cite{18}.

To address the challenges mentioned above, denoising of sequences has garnered increasing attention from researchers. Recent studies demonstrate that using denoising methods in recommender systems can lead to more efficient model training and better performance at a reasonable computational cost \cite{19,20,21}. The existing denoising process involves two steps: recognizing noise and handling noisy sequences.

In practice, recognizing for noise typically judges sequences with high loss values as noisy sequences. Based on the handling of noisy sequences, existing methods can be categorized into two types: truncated denoising and reweighted denoising. For the truncated denoising method \cite{22,23}, the objective is to train a network capable of recognizing noise and discarding noisy sequences, allowing the model to only learn from clean sequences. Regarding the reweighted denoising method \cite{18,24}, once noisy sequences are recognized, this method tends to assign smaller weights to these sequences throughout the entire model training process, thereby reducing the contribution of these sequences to the recommender model.

Although these denoising methods contribute to improving recommender model's performance, user behavior encompasses diverse interest and motivations. Some interactions may be temporary, random or influenced by other factors, which increases the difficulty of recognizing between noisy and clean sequences. Moreover, due to complex data distributions and inherent learning difficulties, high loss values do not necessarily indicate noisy sequences. Additionally, the presence of thresholds in the truncated denoising method heavily relies on the sampling distribution during the decision-making process, inevitably discarding many clean sequences and potentially exhibiting biased selections \cite{18}. Reweighted denoising method requires specific configurations for a given model or recommendation task, which can be time-consuming and challenging to transfer to other settings \cite{25}.

To address the aforementioned issues, from an intuitive perspective, we posit that using a noise recognition model to identify noise sequences and then assigning smaller weights to these sequences to mitigate their influence can enhance the performance of the recommender model. Unlike traditional noise recognition methods, we propose a direct method by constructing a noise recognition model as an auxiliary task to specifically identify noisy sequences. Moreover, to mitigate the impact of reduced training data on the recommender model, we use a novel adaptive reweighting method: training the noise recognition model and the recommender model jointly. This method allows for assigning the most suitable weights for different sequences, optimizing the performance of the recommender model. 

Initially, we construct a noise recognition model to differentiate between clean and noisy sequences in the original dateset. Given the difficulty of identifying noisy sequences within the original data \cite{25}, we artificially create noisy sequences by replacing historical click items of the original sequences with random data. Due to the inherent limitations of human intervention, the artificially replaced noisy sequences may not fully replicate the authentic noisy sequences present in the original sequences. However, since certain authentic noisy sequences also result from users' sporadic, unintentional clicks, there are some similarities between them. Based on the assumption of the existence of certain similarities, we believe that the artificially replaced noisy sequences can represent a portion of the original noisy sequences, thus we regard the artificially replaced noisy sequences as noise data. Given the scarcity of true noise data within the original sequences, we regard the original sequences as clean data. At this point, we can conduct labeled training for the noise recognition model.

Furthermore, we cannot simply discard the noisy sequences from the original sequences, as these noisy sequences may contain factors that are beneficial for the training of the recommender model, and different noisy sequences have varying impacts on the training of the recommender model. Consequently, we use a novel adaptive reweighting method. Taking into account that a fixed weighting strategy does not adapt to model variations and that the contributions of noisy data to model training are not uniform, we opt to design the sequence weights as learnable parameters associated with denoising method and beneficial for the performance of the recommender model. Specifically, we train the noise recognition model using original sequences and randomly replaced noisy sequences. The noise recognition model then weights non-overlapping original sequences not used in its training. These weighted sequences are subsequently used to train the recommender model. This joint training is accomplished through auxiliary task, ensuring that the noise recognition model accurately identifies noisy sequences while optimizing the results of sequence reweighting. 

After training the noise recognition model, the noise recognition model becomes adept at accurately distinguishing between these two types of sequences. In other words, the noise recognition model tends to classify the original sequences it was trained on as clean sequences, which results in the inability to recognize the noisy sequences in the original sequences. Taking this issue into consideration, we choose to use the non-overlapping original sequences that were not involved in the training of the noise recognition model as inputs for the recommender model allows us to determine which of the input sequences used during the training of the recommender model contain noise.

The main contributions of this work are:
\begin{itemize}
\item We introduce a novel self-supervised Auxiliary Task Joint Training (ATJT) method, where the weights obtained from the joint training of the noise recognition model and the recommender model are reweighted onto the sequences used for training the recommender model. This method enhances the performance of the recommender model.
\item The ATJT method is versatile and can be applied to various underlying recommender models.
\item We evaluate the ATJT method on three datasets using a consistent base model. Experimental results show that our method improves recommender model performance.
\end{itemize}

The paper is structured as follows: \secref{section 2} provides a comprehensive overview of related work, focusing on CTR models and denoising methods. \secref{section 3} introduces the preliminary work, describes the training processes for both the noise recognition and recommender models, and explains the ATJT method. \secref{section 4} presents the experimental setup, results and model analysis. \secref{section 5} concludes the paper with a summary of our work and discusses future research directions.

\section{Related Work}\label{section 2}
In this section, we introduce the CTR Models and provide a comprehensive overview of the methods related to sequence denoising in CTR Models.

\subsection{CTR Models}
In recent years, deep learning based models have gained significant traction in CTR prediction \cite{15}. These models exhibit strong representation learning capabilities, enabling them to capture more intricate and challenging patterns and features. Existing deep learning based recommender models can be broadly categorized into two types: sequence-based \cite{26,27,28,29,7,30,31,32,33,34} and graph-based \cite{35,36,37}. We propose a sequence-based denoising method in this paper. Consequently, this subsection focuses on sequence-based recommender models. Wide \& Deep \cite{11} and DCN \cite{14} leverage the memory and generalization capabilities of feature interactions by combining traditional generalized linear models with deep neural networks. DIN \cite{7} uses self-attention mechanisms to enhance the representation of user interest. SASRec \cite{38} and S3Rec \cite{39} utilize multi-head self-attention mechanism to model relationships within sequences. PS-SA\cite{40} employs a learnable progressive sampling strategy to identify the most valuable items. FEARec \cite{41} enhances recommendation by converting user historical behavior sequences into frequency domain representations and combining them with a self-attention mechanism.

CTR models leverage self-supervised learning \cite{42} methods to improve data utilization and learn feature representations. For instance, DuoRec \cite{43} and MPT \cite{44} enhance item embedding distributions through contrastive learning. ICL \cite{45} and simple CL method \cite{46} address data sparsity and popularity bias by learning user intent representations. Pre-training GNN \cite{47}, multi-channel hypergraph convolutional network \cite{48}, DHCN \cite{49} and self-supervised tri-training \cite{50} integrate self-supervised learning with other relevant techniques to enhance the performance of recommender systems.

\subsection{Denoising Methods}
Identifying noisy sequences is an essential step in sequence denoising. DROP \cite{51} and three instance selection methods \cite{52} discuss how to reduce the number of sequences in the training set without affecting classification accuracy. AutoDenoise \cite{25} deletes sequences that have a counteractive effect on the model through rewards. Hierarchical Reinforcement Learning for Course Recommendation in MOOCs \cite{53} removes noisy courses by jointly training of a hierarchical reinforcement learning-based modifier and a basic recommender model. DeCA \cite{24} determines noisy sequences by analyzing the discrepancies in user preferences predicted by two recommender models. MMInfoRec \cite{54} and ContrastVAE \cite{55} address issues such as sparsity and uncertainty in recommender systems by leveraging contrastive learning techniques. DT4SR \cite{56} effectively resolves the problem of neglecting user dynamic preferences and item relationships in traditional methods by introducing uncertainty into sequential modeling. SDK framework \cite{57} deals with the challenges of Knowledge Graphs (KGs) in knowledge-aware recommendation by modeling hyper-relational facts and using self-supervised learning mechanisms. SGL \cite{58} improves the recommendation performance of long-tail items and the robustness against interaction noises by using an auxiliary self-supervised learning task. We propose a denoising auxiliary task neither requires considering the impact on the model nor adds excessive additional training steps. We define a model capable of recognizing noise, thereby enhancing the model's performance. 

After recognizing the noisy sequences, we need to handle these sequences to improve the performance of the recommender model. Existing methods for handling noisy sequences can be classified into two categories: truncated denoising \cite{19,21,18} and reweighted denoising \cite{18}. WBPR \cite{19} and T-CE \cite{18} define thresholds for samples, truncating sequences with loss values higher than the threshold at each iteration. IR \cite{21} modifies labels to train downstream modules for recommendation tasks. In R-CE \cite{18}, smaller weights are assigned to high-loss sequences to prevent the model from fitting them too quickly. However, truncated denoising risks filtering out many clean sequences, while reweighted denoising suffers from limited transferability. We propose an ATJT method similar to reweighted denoising, but it addresses limitations by adaptively adjusting the weighting degree.

\section{Methodology}\label{section 3}
In this section, we will introduce the preliminary work and discuss the training processes for both the noise recognition model and the recommender model. We will also provide a detailed explanation of how to implement the ATJT method.

\subsection{Preliminary}\label{section 3.1}
In this paper, we use batch $b$ composed of training sequences as the input for both the noise recognition model and the recommender model. Each batch has a size M, and the sequences have a length N. 

All batches are divided into two groups, $\mathcal{B}^{R}$ and $\mathcal{B}^{D}$. The batch in the first group, denoted as $b_{i}^{R} = \left \{ s_{i,1 },\cdots,s_{i,m},\cdots , s_{i,M} \right \} \in \mathcal{B}^{R}$, undergoes obtaining the weights of historical interaction sequences through the noise recognition model. We then use the reweighted sequences to train the recommender model. We use $s_{i}$ to represent the sequences of the $i$-th batch that are used for training the recommender model. And $\mathcal{B}^{R}$ consists of a total of I batches. The batch in the second group, denoted as $b_{j}^{D} = \left \{ s_{j,1},\cdots,s_{j,m},\cdots , s_{j,M} \right \}\in \mathcal{B}^{D}$, is used to train the noise recognition model capable of accurately recognizing noisy sequences. We use $s_{j}$ to represent the sequences of the $j$-th batch that are used for training the noise recognition model. And $\mathcal{B}^{D}$ consists of a total of J batches. In summary, $\mathcal{B}^{R} \cup \mathcal{B}^{D}=\mathcal{B}$ and $\mathcal{B}^{R} \cap \mathcal{B}^{D} =\phi$.

We further divide the batch $b_{j}^{D}$ into two batches, $b_{j}^{D_{(+)}}$ and $b_{j}^{D_{(-)}}$. $b_{j}^{D_{(+)}}$ represents clean batch within $b_{j}^{D}$ consisting of original sequences. $b_{j}^{D_{(-)}} = \left \{ s_{j,1 }^{'},\cdots,s_{j,m}^{'},\cdots , s_{j,M}^{'} \right \}$ represents noisy batch consisting of randomly replaced sequences from $b_{j}^{D}$, where $s_{j,m}^{'}= \left \{ v_{1},\cdots,v_{n}^{'},\cdots ,v_{N} \right \}$ represents the $m$-th noisy sequence that has undergone random replacement in the $j$-th batch of the noise recognition model. Within the sequence $s_{j,m}^{'}$, $v_{n}^{'}$ represents the $n$-th interaction item that has been randomly replaced. At this point, the second batch transforms into $b_{j}^{D} = \left \{s_{j,1},\cdots,s_{j,m}^{'},\cdots , s_{j,M}\right \}\in \mathcal{B}^{D}$. In summary, $b_{j}^{D_{(+)}} \cup b_{j}^{D_{(-)}}=b_{j}^{D}$ and $b_{j}^{D_{(+)}} \cap b_{j}^{D_{(-)}} =\phi$.

\begin{figure*}[!htb]
\centering
  \includegraphics[width=0.9\linewidth]{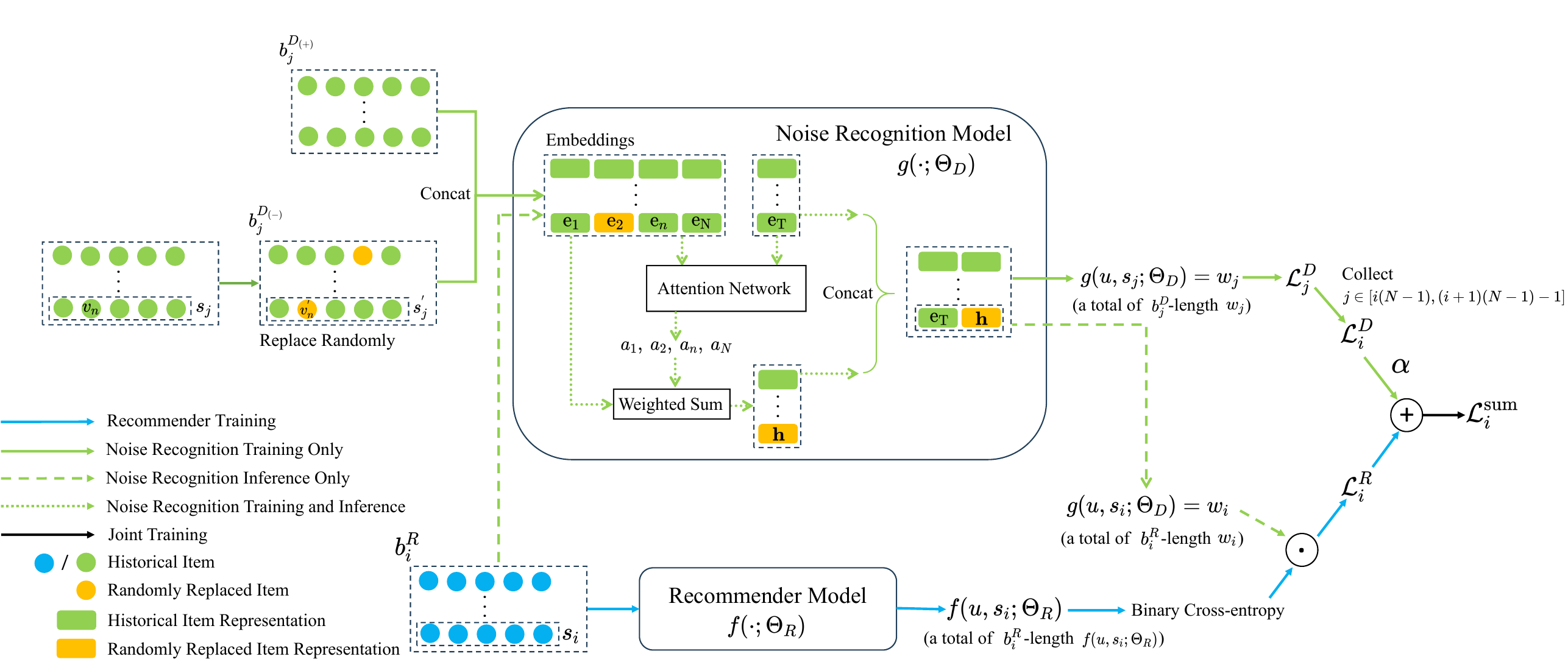}
  \caption{The ATJT method consists of two main components: 1. Training the noise recognition model (compose with Noise Recognition Training Only and Noise Recognition Training and Inference). 2. Training the recommender model using reweighted sequences (compose with Recommender Training, Noise Recognition Inference Only and Noise Recognition Training and Inference).}
  \label{fig_1}
\end{figure*}

We use $f(\cdot ;\Theta _{R})$ to represent the recommender model and $g(\cdot;\Theta _{D})$ to represent the noise recognition model. Given the users' historical interaction sequences $(u,s_{i})\in b_{i}^{R}$, the recommender model can predict the probabilities $f(u,s_{i};\Theta _{R})$ of clicks. Similarly, given $(u,s_{j})\in b_{j}^{D}$, the noise recognition model can predict the probabilities $g(u,s_{j};\Theta _{D})$ of noise contamination.

In summary, we enhance the recommender model's performance by obtaining accurate weights $w_{i}$ for the sequences $s_{i}$ from the joint training of $f(\cdot ;\Theta _{R})$ and $g(\cdot;\Theta _{D})$.

\subsection{Recommender Model}
In the training process of the recommender model, as shown in \figref{fig_1}, given a batch $b_{i}^{R} \in \mathcal{B}^{R}$, the sequences $s_{i}$ in the batch initially pass through the noise recognition model to obtain weights $w_{i}$. Subsequently, the reweighted sequences are used to train the parameters of the recommender model. It is worth noting that the recommender model can be chosen based on specific requirements, such as DIN or DCN. Its training process aligns with these base models.

The CTR prediction of the recommender model can be viewed as a supervised binary classification task. Therefore, we optimize the recommender model using a binary cross-entropy loss function. Additionally, considering the impact of noisy sequences on the training of the recommender model, it is essential to recognize and assign smaller weights to mitigate the influence of noisy sequences. Consequently, we define the loss function for the recommender model as follows:
\begin{equation}
\begin{split}
\mathcal{L}_{i}^{R} &= -\frac{1}{|b_{i}^{R}|}\sum_{s_{i} \in b_{i}^{R}}^{|b_{i}^{R}|}(w_{i}(y_{i}\log f(u,s_{i};\Theta _{R}) \\
& \quad + (1-y_{i})\log(1-f(u,s_{i};\Theta _{R})))),
\end{split}
\label{eqn_1}
\end{equation}
where $y_{i}$ and $f(u,s_{i};\Theta _{R})$ represent the labels for clicks and the predicted probabilities of clicks for the sequences $s_{i}$ in $b_{i}^{R}$, respectively. $w_{i}$ represents the weights of the sequences $s_{i}$. Typically, noisy sequences have smaller weights $w_{i}$ compared with clean sequences in model training (as shown in our experiments in \subsecref{section 4.2.4}). This approach reduces the impact of noisy sequences on model performance \cite{18,24}. We will elaborate on how to determine the sequence weights $w_{i}$ that improve the performance of the recommender model in \subsecref{section 3.3.2} and \subsecref{section 3.4.2}.

\begin{figure*}[!htb]
\centering
  \subfigure[]{
    \includegraphics[width=5.5cm,height=2.5cm]{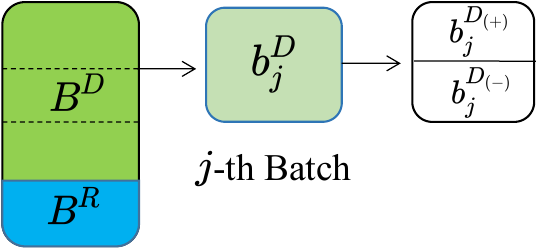}}
  \hspace{1.5cm}
  \subfigure[]{
    \includegraphics[width=7cm,height=2.5cm]{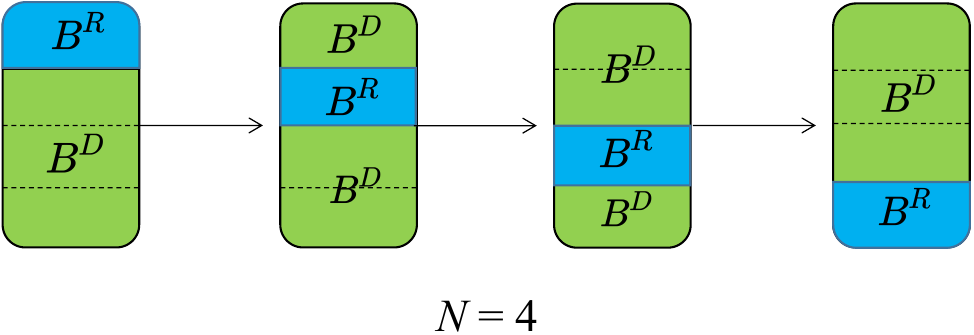}}
  \caption{The left figure shows the division of noisy and clean sequences within a batch in the noise recognition model, following a 1:1 ratio. The right figure illustrates the partitioning of training data for the recommender model and the noise recognition model (blue represents training data $\mathcal{B}^{R}$ for the recommender model, and green represents training data $\mathcal{B}^{D}$ for the noise recognition model). (a) Data Replacement. (b) Data Partition.}
  \label{fig_2}
\end{figure*}

\subsection{Noise Recognition Model}
To build a noise recognition model capable of accurately distinguishing noisy sequences from clean sequences and weighting the sequences for the recommender model, we opt for a self-supervised training method. In this subsection, we will focus on two essential components: data replacement and weight generation.

\subsubsection{Data Replacement}\label{section 3.3.1}
As shown in \figref{fig_2}(a), we use the batch $b_{j}^{D}= b_{j}^{D_{(+)}} \cup b_{j}^{D_{(-)}}$ as the input for the noise recognition model, where $b_{j}^{D_{(+)}}$ represents clean batch consisting of original sequences, labeled as 1. And $b_{j}^{D_{(-)}}$ represents noisy batch composed of randomly replaced noisy sequences, labeled as 0. While the selection of $b_{j}^{D_{(-)}}$ from $b_{j}^{D}$ is not fixed, it should not be too scant. Specifically, we assume that there are very few noisy sequences in $b_{j}^{D_{(+)}}$. If we select too few sequences in $b_{j}^{D_{(-)}}$, it may lead to a situation where the extremely few noisy sequences in $b_{j}^{D_{(+)}}$ outnumber the sequences in $b_{j}^{D_{(-)}}$, meaning that the number of sequences in $b_{j}^{D_{(+)}}$ labeled as 1 while actually being 0 is greater than the number of sequences in $b_{j}^{D_{(-)}}$ labeled as 0. This situation could lead the noise recognition model to incorrectly learn noisy sequences as positive (labeled 1). Hence, it is essential to ensure an adequate number of sequences in $b_{j}^{D_{(-)}}$ to avoid an unstable situation that could lead the noise recognition model to learn in the wrong direction. Up to this point, we have discussed the training method for the recommender model and how input sequences for the noise recognition model are generated.

\subsubsection{Weight Generation}\label{section 3.3.2}
In this subsection, we will describe the method for generating weights $w_{i}$. As depicted in \figref{fig_1}, the noise recognition model is a sequence-to-value model. The model takes $b_{j}^{D} \in \mathcal{B}^{D}$ as input. For $s_{j} \in b_{j}^{D}$, where $s_{j}$ consists of items with length $N$ and a target item to be predicted. $s_{j}$ first passes through the neural network's embedding layer, it is transformed into the sequences of embeddings $\left[\mathbf{e}_{1},\cdots,\mathbf{e}_{n},\cdots,\mathbf{e}_{N},\mathbf{e}_{\text{T}}\right]$, in which $\mathbf{e}_{n}$ represents the embedding of the $n$-th item in the sequences $s_{j}$ after passes through the neural network's embedding layer. Then, we pass it through the attention network to obtain user hidden representation of the sequences $s_{j}$:
\begin{equation*}
\mathbf{h} = \sum_{n = 1}^{N} a_n \mathbf{e}_{n},
\end{equation*}
where 
\begin{equation*}
a_n = \frac{\text{MLP}(\mathbf{e}_{n}||\mathbf{e}_{\text{T} })}{\sum^{N}_{n'= 1}\text{MLP}(\mathbf{e}_{n'}||\mathbf{e}_{\text{T} })}.\\
\end{equation*}
$||$ represents the concatenation of embeddings. Subsequently, we concat $\mathbf{h}$ with the embedding $\mathbf{e}_{\text{T}}$ of the target item, and then pass the results through a MLP to produce the weights $w_{j}$. Given that our noise recognition method can be viewed as a self-supervised binary classification task, we use the binary cross-entropy loss function for optimization:
\begin{equation}
\begin{split}
\mathcal{L}_{j}^{D} &= -\frac{1}{|b_{j}^{D}|}\sum_{s_{j} \in b_{j}^{D}}^{|b_{j}^{D}|}(y_{j}\log g(u,s_{j};\Theta _{D}) \\
& \quad + (1-y_{j}) \log(1-g(u,s_{j};\Theta _{D}))),
\end{split}
\label{eqn_2}
\end{equation}
where $y_{j}$ and $g(u,s_{j};\Theta _{D})=w_{j}$ represent the labels for noise and the predicted probabilities of noise for the sequences $s_{j} $ in $b_{j}^{D}$, respectively.

\begin{figure*}
    \centering
    \includegraphics[width=0.8\linewidth]{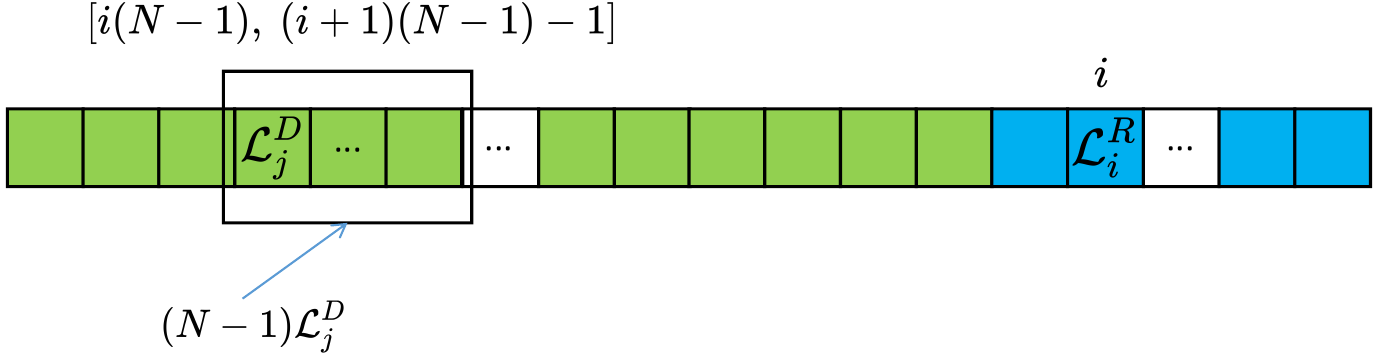}
    \caption{When training the recommender model with the $i$-th set of data from $\mathcal{B}^{R}$, concurrently train the noise recognition model with data from $\mathcal{B}^{D}$ in the range $[i(N-1), (i+1)(N-1)-1]$ (green represents training data for the noise recognition model, and blue represents training data $\mathcal{B}^{R}$ for the recommender model).}
    \label{fig_6}
\end{figure*}

The noise recognition model similarly uses training set $b_{i}^{R} \in \mathcal{B}^{R}$, which is used in training the recommender model, as input. This set of sequences is non-overlapping with the training sequences used for the noise recognition model, which will be explained in detail in \subsecref{section 3.4.1}. At this point, we can use the results $g(u,s_{i};\Theta _{D})$ output by the noise recognition model as the weights for the training sequences of the recommender model, namely the weights $w_{i}$ in \equref{eqn_1}.

Furthermore, the noise recognition model is used only during the training phase to help the recommender model learn better parameters. It is not used during the evaluation phase. Therefore, the ATJT method does not increase the number of parameters in the recommender model.

\subsection{ATJT Method}

\subsubsection{Data Partition.}\label{section 3.4.1}
To fully use the data, after fitting the parameters of both the recommender model and the noise recognition model, we can reverse the training data for $\mathcal{B}^{R}$ and $\mathcal{B}^{D}$. This means that $B^{D}$ optimizes the recommender model while $\mathcal{B}^{R}$ optimizes the noise recognition model. It is worth noting that the recommender model continues to use the original model in the subsequent training, while the noise recognition model is trained using a duplicate model. The purpose of this is to ensure that all data can be used to train the recommender model, while the noise recognition model does not fit all the data. Throughout the training process, the reweighted recommender model and the noise recognition model are trained together. This ensures that the noise recognition model can accurately recognize noisy sequences while optimizing the results of sequence reweighting.

Two important points to note are: first, we only use the original sequences to train the recommender model, and the noise recognition model is trained with original sequences and randomly replaced noisy sequences. Second, if there is a need to make more extensive use of the data, the training set sequences can be divided into \textit{N} groups instead of two groups. As shown \figref{fig_2}(b), we demonstrate a training method where the training set is divided into four groups. In extreme cases, only one sequence receives the best reweighting output by the noise recognition model and trains the recommender model, while the rest of the sequences are input into the noise recognition model to achieve the best recognition performance in training. However, this method increases the number of duplicate models for training the noise recognition model. Therefore, the minimum grouping is two groups, and the maximum grouping is the size of the training set sequences $|\mathcal{B}^{R} \cup \mathcal{B}^{D}|$. The specific grouping can be chosen based on available resources and performance considerations.

\subsubsection{Loss Function}\label{section 3.4.2}

\begin{figure*}[ht]
    \centering
    \begin{minipage}{\textwidth}
\begin{algorithm}[H]
	\caption{Overall Optimization Algorithm of Model Training}\label{algo1}
	\begin{algorithmic}[1]
		\REQUIRE $b_{i}^{R} \in \mathcal{B}^{R}$, $b_{j}^{D} \in \mathcal{B}^{D}$, recommender model $j(\cdot ;\Theta _{R} )$, noise recognition model $g(\cdot;\Theta _{D})$, dataset groups \textit{N}, reweighted recommender model loss $\mathcal{L}_{i}^{R}$, noise recognition model loss $\mathcal{L}_{j}^{D}$, (\textit{N}-1) batches of noise recognition model loss $\mathcal{L}_{i}^{D}$, sum loss $\mathcal{L}_{i}^{\text{sum}}$
		\ENSURE trained recommender model $f^{*}(\cdot ;\Theta _{R})$
		
		\STATE Create $\mathcal{L}_{i}^{R}$, $\mathcal{L}_{j}^{D}$, $\mathcal{L}_{i}^{\text{sum}}$
		\STATE Initialize $f(\cdot;\Theta _{R})$
		\FOR{$n \in [1,N]$}
			\STATE Initialize $g(\cdot;\Theta _{D})$
			\FOR{$i \in [1,|\mathcal{B}^{R}|]$}
				\STATE Estimate $w_{i}$ using noise recognition model $g(\cdot;\Theta _{D})$
				\STATE Get $\mathcal{L}_{i}^{R}$ by \equref{eqn_1}
				\STATE $\mathcal{L}_{i}^{D} \gets 0$
				\FOR{$j \in [i(N-1), (i+1)(N-1)-1]$}
					\STATE $b_{j}^{D} = b_{j}^{D_{(+)} } \cup b_{j}^{D_{(-)} }$, $|b_{j}^{D_{(+)} }| = |b_{j}^{D_{(-)}}|$
					\STATE $s_{j}^{'} = [v_{1}, \cdots, v_{n}^{'}, \cdots, v_{N}] \in b_{j}^{D_{(-)}}$ represents replaced sequence as described in \subsecref{section 3.1}
					\STATE Get $\mathcal{L}_{j}^{D}$ by \equref{eqn_2}
					\STATE $\mathcal{L}_{i}^{D} \mathrel{+}= \mathcal{L}_{j}^{D}$
				\ENDFOR
				\STATE $\mathcal{L}_{i}^{\text{sum}} \gets \mathcal{L}_{i}^{R} + \alpha \frac{1}{N-1} \mathcal{L}_{i}^{D}$
				\STATE Update $\Theta _{R}$
				\STATE Update $\Theta _{D}$
			\ENDFOR
		\ENDFOR
	\end{algorithmic}
\end{algorithm}
    \end{minipage}
\end{figure*}

In this subsection, we focus on how to jointly train the recommender model with the noise recognition model by computing the loss value. When we partition the training set sequences into \textit{N} groups, due to $|\mathcal{B}^{R}|:|\mathcal{B}^{D}|=1:N-1$, the batches used for training the recommender model should be in a ratio of $1:N-1$ compared with those used for training the noise recognition model, as illustrated in \figref{fig_6}. Therefore, the loss function for the noise recognition model should be:
\begin{equation*}
\mathcal{L}_{i}^{D}=\frac{1}{N-1}\sum_{j=i(N-1)}^{(i+1)(N-1)-1}\mathcal{L}_{j}^{D},
\end{equation*}
where $i$ represents the index of the batch used by the recommender model. $N-1$ represents the number of batches used by the noise recognition model corresponding to one batch of data used for training the recommender model, and $\mathcal{L}_{j}^{D}$ represents the binary cross-entropy loss function when training the noise recognition model with the $j$-th batch of data.
During the training process, we combine the loss of the recommender model with the loss of the noise recognition model using a scaling factor $\alpha$ to obtain the total loss for model training:
\begin{equation*}
\mathcal{L}_{i}^{\text{sum}}=\mathcal{L}_{i}^{R}+\alpha\mathcal{L}_{i}^{D},
\end{equation*}
where $\alpha$ represents tunable parameters that allows us to control the learning rates of the recommender model and the noise recognition model, thereby achieving the goal of joint training. Joint training enables the noise recognition model to learn the weights $w_{i}$ for sequences $s_{i}$ in $b_{i} ^{R}$ (as defined in \equref{eqn_1}), which more suitable for training the recommender model while accurately recognizing noise. This enables the recommender model to achieve better performance with the weighted sequences.

\subsubsection{Overall Optimization Algorithm of Model Training}
The joint training process is illustrated in Algorithm \ref{algo1}. The joint training consists of two parts: the calculation of the $\mathcal{L}_{i}^{R}$ for the recommender model (lines 6-7) and the calculation of the $\mathcal{L}_{i}^{D}$ for the noise recognition model (lines 8-14). We achieve joint training by summing the loss values from these two parts (line 15). Specifically, we start by initializing the recommender model $f(\cdot ;\Theta _{R})$ (line 2). Next, we iterate over all groups in the training set as described in \subsecref{section 3.4.1} (line 3). We then initialize the noise recognition model (line 4) and retrieve the $i$-th batch from the recommender model training set (line 5). The sequences from the $i$-th batch are passed through the noise recognition model $g(\cdot;\Theta _{D})$ to determine their weights $w_{i}$ (line 6). These sequences are then input into the recommender model, and the weighted loss is calculated. After averaging the loss for all sequences in the batch, we obtain $\mathcal{L}_{i}^{R}$ (line 7). Subsequently, we iterate over the batches from the ($i(N-1)$)-th to the ($(i+1)(N-1)-1$)-th in the noise recognition model training set (line 9). From the batch of the noise recognition model, we select half of the sequences. For these sequences, we perform random replacements of items, considering them as noisy sequences (line 10 and 11). The original sequences and the noisy sequences from set $b_{j}^{D}$ are input into the noise recognition model, and the loss is calculated. After averaging the loss for all sequences in the batch, we obtain $\mathcal{L}_{j}^{D}$ (line 12). Next, we accumulate all $\mathcal{L}_{j}^{D}$ values during the iteration onto $\mathcal{L}_{i}^{D}$ to obtain the final noise recognition model loss (line 13). Finally, we add $\mathcal{L}_{i}^{R}$ and $\mathcal{L}_{i}^{D}$ to obtain $\mathcal{L}_{i}^{\text{sum}}$ (line 15). At this point, we can jointly optimize the recommender model and the noise recognition model based on $\mathcal{L}_{i}^{\text{sum}}$ (line 16 and 17). At this point, we have completed the training of the noise recognition model and explained how to implement the ATJT method.

\section{Experiments}\label{section 4}

We conduct extensive experiments to address the following three questions:

\begin{itemize}
\item RQ1: How does the performance of the ATJT method compare with the base model?
\item RQ2: What is the impact of different types of noisy sequences generation on performance?
\item RQ3: How does different sequence weighting training methods affect performance?
\end{itemize}

\subsection{Experimental Setting}

\subsubsection{Datasets and Baselines}
We evaluate our method using the MovieLens20M\jz{\textcolor{blue}{1}}{\href{https://grouplens.org/datasets/movielens/20m}{https://grouplens.org/datasets/movielens/20m}, Jun. 2024.}, Amazon (Electro)\jz{\textcolor{blue}{2}}{\href{http://jmcauley.ucsd.edu/data/amazon/}{http://jmcauley.ucsd.edu/data/amazon/}, Jun. 2024.} and Yelp\jz{\textcolor{blue}{3}}{\href{https://www.kaggle.com/datasets/yelp-dataset/yelp-dataset/data}{https://www.kaggle.com/datasets/yelp-dataset/yelp-dataset/data}, Jun. 2024.} datasets. We select these three datasets for two reasons: 1) They represent diverse scenarios, namely an online movie platform and an e-commerce platform, with varying levels of product diversity. 2) They differ in size and characteristics. The statistical data for MovieLens20M, Amazon (Electro) and Yelp are shown in \tableref{label_1}.

\tabcolsep 12pt
\renewcommand\arraystretch{1.3}
\setlength{\tabcolsep}{8pt}
\begin{table}[H]
\centering 
\captionof{table}{\label{label_1} Statistical information for datasets} 
\vspace{-2mm}
\footnotesize{
\begin{tabular*}{\linewidth}{cccc}
\hline\hline\hline
    Dataset & \#User & \#Item & \#Sample\\
\hline
    MovieLens20M & 138,493 & 27,278 & 20,000,263\\
    Amazon (Electro) & 192,403 & 63,001 & 1,689,188\\
    Yelp & 1,987,929 & 150,346 & 6,990,280\\
\hline\hline\hline
\end{tabular*}
\\\vspace{1mm}\parbox{8.3cm}{Note: \# represents Number of.}
}
\end{table}

The MovieLens20M and Amazon (Electro) datasets all consist of features such as user ID, historical interaction item IDs and their corresponding categories. In Yelp dataset, each item has features including its business\_id, city, postal\_code, star rating and categories. Each user has features including the user\_id, useful, funny, cool and average star rating.

We employ several advanced recommender models as base models, including Wide \& Deep \cite{11}, DCN \cite{14}, DIN \cite{7}, SASRec \cite{38}, S3Rec \cite{39} and FEARec \cite{41}. We use the results of these six base models on three datasets as the baseline and compare them with the ATJT method.

In summary, we conduct a total of 6 (the number of recommender models) * 2 (the number of contrastive models) experiments to assess the performance improvement of the ATJT method on three specified datasets for the recommender model.

\subsubsection{Evaluation Protocol}

\renewcommand\arraystretch{1.3}
\begin{table*}[!htb]
\centering
\caption{\label{label_2} Experimental results on three datasets based on different recommender models}
\vspace{-2mm}
{\footnotesize
\begin{tabular*}{\linewidth}{@{\extracolsep{\fill}}cccccccccc}
\hline\hline\hline
    Model & \multicolumn{3}{c}{MovieLens20M} & \multicolumn{3}{c}{Amazon (Electro)} & 
    \multicolumn{3}{c}{Yelp} \\ 
    \cmidrule(lr){2-4} \cmidrule(lr){5-7} \cmidrule(lr){8-10} 
    & AUC & HR@5 & NDCG@5 & AUC & HR@5 & NDCG@5 & AUC & HR@5 & NDCG@5 \\
    \hline
    Wide \& Deep & 0.8217 & 0.5052 & 0.1117 & 0.8535 & 0.5685 & 0.1224 & 0.7503 & 0.3866 & 0.0859 \\
    Wide \& Deep+ATJT & 0.8310 & 0.5124 & 0.1143 & 0.8572 & 0.5743 & 0.1248 & 0.7599 & 0.3974 & 0.0897 \\
    +RI & 1.13\% & 1.43\% & 2.33\% & 0.43\% & 1.02\% & 1.96\% & 1.28\% & 2.79\% & 4.42\% \\
    \hline
    DCN & 0.8438 & 0.5091 & 0.1145 & 0.8729 & 0.5865 & 0.1376 & 0.7881 & 0.3823 & 0.0912 \\
    DCN+ATJT & 0.8450 & 0.5137 & 0.1157 & 0.8735 & 0.5879 & 0.1382 & 0.7885 & 0.3917 & 0.0927 \\
    +RI & 0.14\% & 0.90\% & 1.05\% & 0.06\% & 0.24\% & 0.44\% & 0.05\% & 2.46\% & 1.64\% \\
    \hline
    DIN & 0.8516& 0.5222 & 0.1166 & 0.8748 & 0.6011 & 0.1406 & 0.8032 & 0.4506 & 0.1012 \\
    DIN+ATJT & 0.8519 & 0.5226 & 0.1167 & 0.8749 & 0.6024 & 0.1411 & 0.8036 & 0.4514 & 0.1015 \\
    +RI & 0.04\% & 0.08\% & 0.09\% & 0.01\% & 0.22\% & 0.36\% & 0.05\% & 0.18\% & 0.30\% \\
    \hline
    SASRec & 0.8475 & 0.5222 & 0.1180 & 0.8767 & 0.5881 & 0.1368 & 0.7747 & 0.4109 & 0.0912 \\
    SASRec+ATJT & 0.8490 & 0.5224 & 0.1183 & 0.8772 & 0.5889 & 0.1370 & 0.7763 & 0.4125 & 0.0917 \\
    +RI & 0.18\% & 0.04\% & 0.25\% & 0.06\% & 0.14\% & 0.15\% & 0.21\% & 0.39\% & 0.55\% \\
    \hline
    S3Rec & 0.8490 & 0.5311 & 0.1170 & 0.8784 & 0.6035 & 0.1420 & 0.8033 & 0.4438 & 0.0999 \\
    S3Rec+ATJT & 0.8499 & 0.5312 & 0.1172 & 0.8787 & 0.6058 & 0.1424 & 0.8048 & 0.4460 & 0.1003 \\
    +RI & 0.11\% & 0.02\% & 0.17\% & 0.03\% & 0.38\% & 0.28\% & 0.19\% & 0.50\% & 0.40\% \\
    \hline
    FEARec & 0.8537 & 0.5311 & 0.1208 & 0.8804 & 0.6063 & 0.1431 & 0.8049 & 0.4512 & 0.1022 \\
    FEARec+ATJT & 0.8539 & 0.5331 & 0.1213 & 0.8812 & 0.6094 & 0.1438 & 0.8060 & 0.4566 & 0.1033 \\
    +RI & 0.02\% & 0.38\% & 0.41\% & 0.09\% & 0.51\% & 0.49\% & 0.14\% & 1.20\% & 1.08\% \\
\hline\hline\hline
\end{tabular*}
}
\end{table*}

To accurately assess the performance of the recommender model, we first divide users' historical interaction sequences into training and testing sets in a 4:1 ratio. In this setup, we use the training set to train both the noise recognition model and the recommender model, while the testing set is used to evaluate the performance of the recommender model. Notably, we need to ensure that users' historical interaction sequences in the training and testing sets are non-overlapping. Additionally, to avoid the issue described in \subsecref{section 3.4.1}, where the noise recognition model fits the training data, we also need to ensure that the historical interaction sequences used for training the recommender model and the noise recognition model are non-overlapping.

We evaluate the testing set using standard AUC (Relative Improvement Area Under the ROC Curve) scores, HR@5 and NDCG@5. These three metrics are widely used in click prediction tasks \cite{7,8}. Higher values for all three metrics indicate superior model performance.

\subsubsection{Implementation Details}\label{section 4.1.3}
The construction of the ATJT method is based on the PyTorch framework. We encapsulate the noise recognition model into a class, allowing it to be integrated as a plugin with most recommender models. The implementation of the noise recognition model follows a unified structure when integrated with different underlying recommender models. The implementation of the noise recognition model relies on an attention mechanism. Specifically, we implement it as an attention model with embedding and output layers. The attention part consists of two layers of MLPs. For MLP (1), we set the linear layers as (64, 32), and for MLP (2), we set the linear layers as (32, 1). Each MLP consists of a linear layer, a PReLU activation function and a dropout operation (rate=0.5). The output part after SUM pooling consists of three layers of MLPs. For MLP (1), we set the linear layers as (40, 256), for MLP (2), we set the linear layers as (256, 64), and for MLP (3), we set the linear layers as (64, 1). The structure of MLPs is identical to the attention part. The output layer is implemented with a sigmoid function, with a dimension of 1, in order to obtain different weights for training the recommender model with noisy and clean sequences. The noise recognition model is uniformly optimized using the Adagrad optimizer, and a learning rate search is conducted from \{0.1, 0.01, 0.001, 0.0001\}.

When training recommender models, each method follows the following steps: 1) When training DCN and Wide \& Deep models, we treat historical interactions as item features. The DNN architectures for DCN and Wide \& Deep are set as (128, 128) and (256, 128), respectively. 2) For DIN model, we use user ID, historical interaction item IDs and their corresponding categories as input features, following \cite{7}. 3) For SASRec and S3Rec models, we use historical interaction item IDs as input features, following \cite{38} and \cite{39}. 4) When training the FEARec model, our input features are the same as those used in the DIN model. Additionally, we use the default hyperparameter configurations provided by the original author of the model on GitHub\jz{\textcolor{blue}{4}}{\href{https://github.com/sudaada/FEARec}{https://github.com/sudaada/FEARec}, Jun. 2024}.

\begin{figure*}[!htb]
\centering
  \subfigure[]{
    \includegraphics[width=5cm,height=3.5cm]{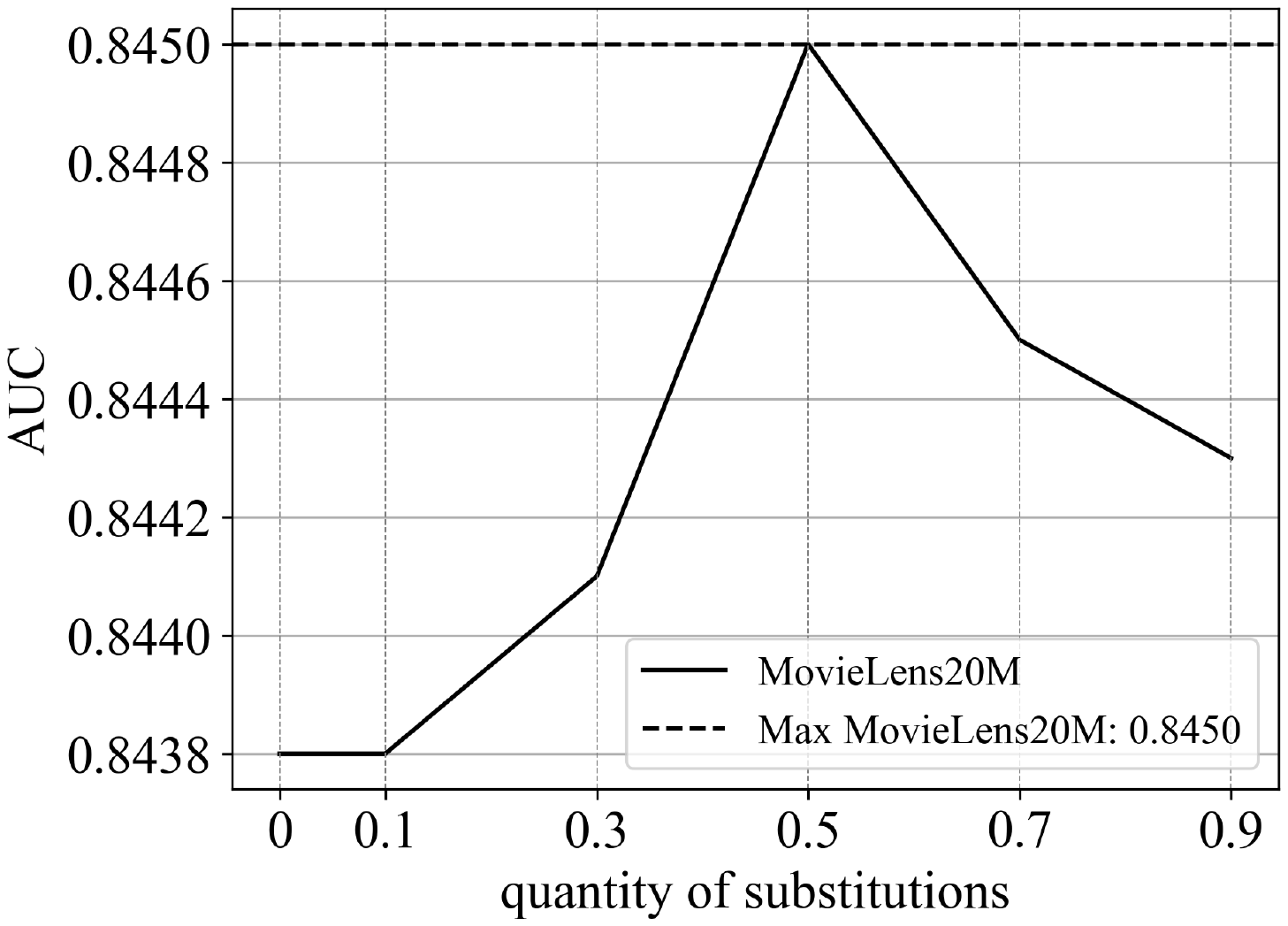}}
    \hspace{0.5cm}
  \subfigure[]{
    \includegraphics[width=5cm,height=3.5cm]{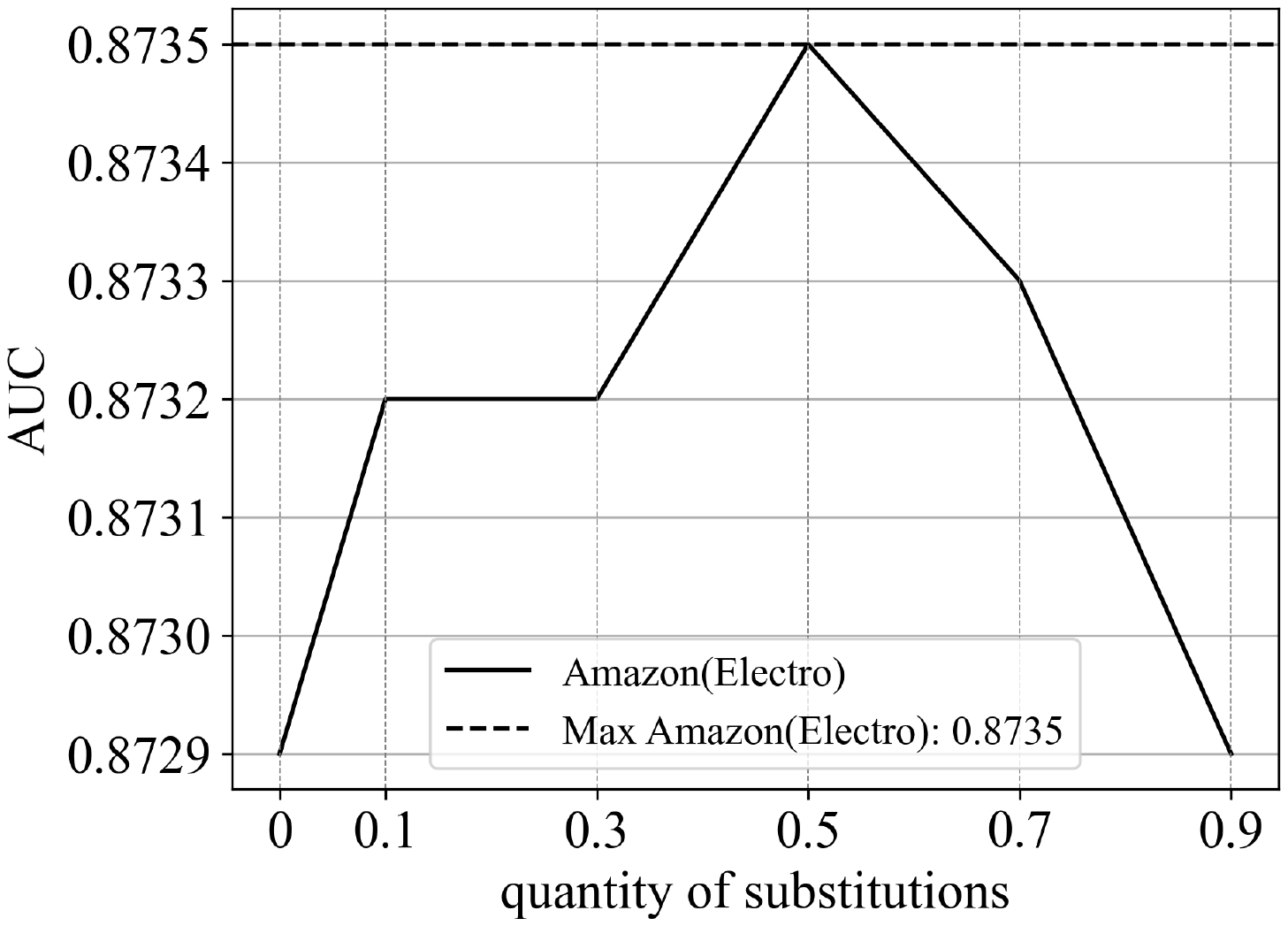}}
    \hspace{0.5cm}
  \subfigure[]{
    \includegraphics[width=5cm,height=3.5cm]{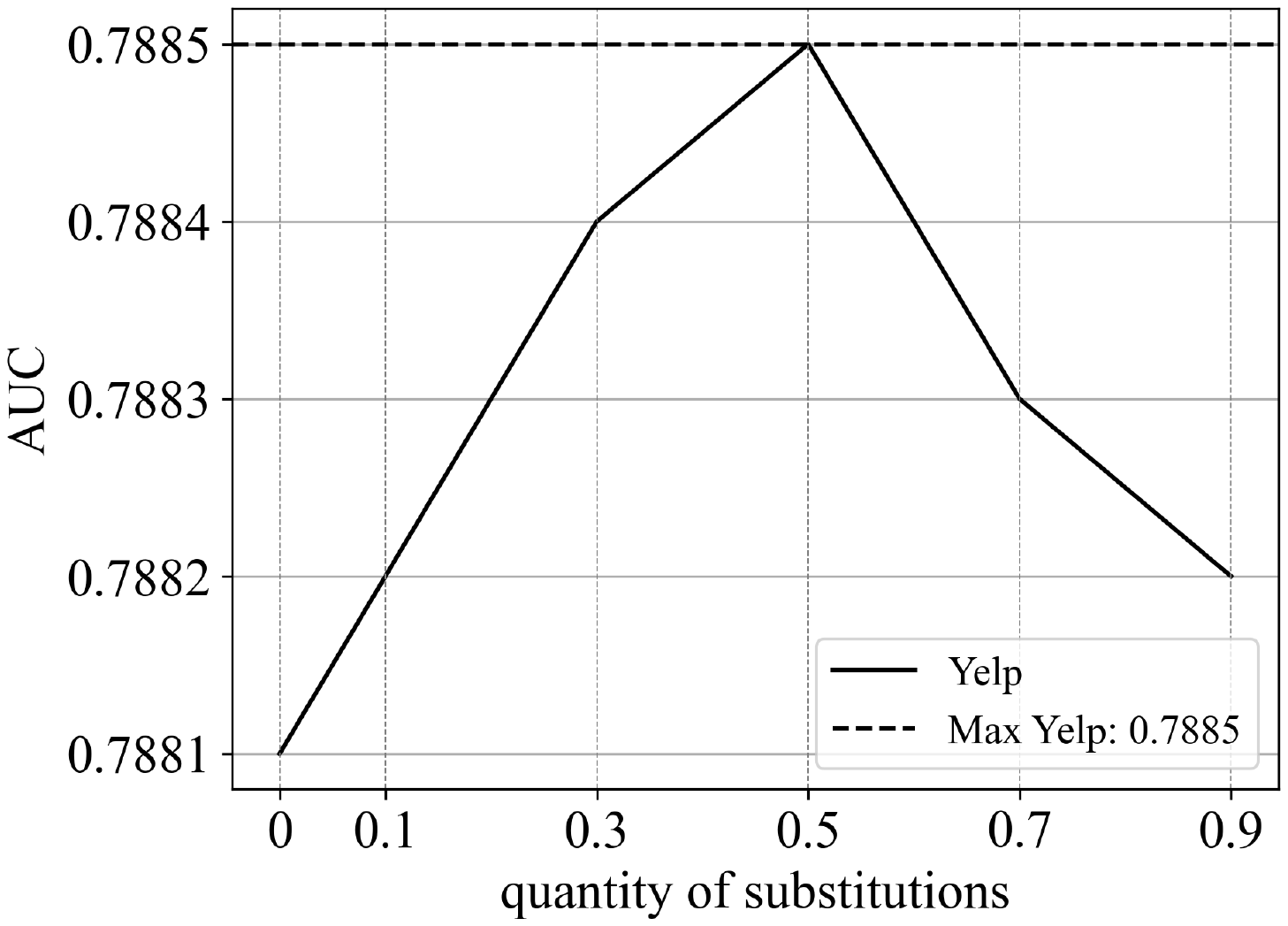}}
  \caption{Analyzing the impact of using different number of sequences as noisy sequences in the training of noise recognition models (Base Model: DCN). (a) MovieLens20M. (b) Amazon (Electro). (c) Yelp.}
  \label{fig_3}
\end{figure*}

\subsection{Experimental Results}

\subsubsection{Overall Performance (RQ1)}
We propose an ATJT method based on six fundamental recommender models and compare their performance with the base recommender models on three different datasets. Experimental results show that the ATJT method outperforms base models in terms of AUC, HR@5 and NDCG@5, as shown in \tableref{label_2}.

We find that the ATJT method yields better improvements in DCN and Wide \& Deep models compared with DIN, SASRec, S3Rec and FEARec models. This phenomenon can be attributed to the attention mechanism possessed by DIN, SASRec, S3Rec and FEARec, which adaptively learns users' interest representations from the historical interaction sequences, thus mitigates the impact of behaviors unrelated to users' interest representations \cite{7}. Furthermore, the enhancement of the ATJT method is more pronounced in the Wide \& Deep model than in the DCN model. The reason may lie in that when we process the input features of the DCN model, we regard historical interactions as item features. Therefore, the cross network captures feature interactions, mitigates the impact of irrelevant features on model performance during training \cite{14}. The Wide \& Deep model lacks attention mechanisms like DIN, multi-head attention mechanism like SASRec, S3Rec and FEARec or the cross network like the DCN model, which can filter out irrelevant or negative behaviors on model optimization. The ATJT method compensates for the Wide \& Deep model's inability to filter out irrelevant or negative behaviors within user actions, resulting in a more noticeable performance improvement.

The ATJT method demonstrates superior performance on the Yelp dataset compared with the MovieLens20M and Amazon (Electro) datasets. This observation may be attributed to the relatively larger size of the Yelp dataset, which provides more data for training more complex models after denoising. The MovieLens20M dataset is also substantial in size, whereas the Amazon (Electro) dataset is relatively smaller, potentially impacting the model's performance after denoising. However, the ATJT method exhibits significant improvements across various base models and data sizes. Furthermore, experiments conducted with six different recommender models indicate the adaptability and efficacy of the ATJT method.

\subsubsection{Noise Generation Analysis (RQ2)}
In this subsection, we analyze the effect of selecting how many sequences in the noise recognition model training data to replace, and the effect of replacing a different number of historical click items in the artificially replaced noisy sequences. \figref{fig_3} shows the impact of selecting 0.1, 0.3, 0.5, 0.7 and 0.9 of the data in the noise recognition model training as noisy sequences on the performance of the recommender model. \tableref{label_3} shows the impact of replacing 1, 2, 3, 5 and 10 historical click items in the artificially replaced noisy sequences on the performance of the recommender model.

\tabcolsep 12pt
\renewcommand\arraystretch{1.3}
\setlength{\tabcolsep}{8pt}
\begin{table}[H]
    \centering 
    \captionof{table}{\label{label_3} Analyzing the impact of replacing 1, 2, 3, 5 and 10 historical click items in the artificially replaced noisy sequences on the performance of the recommender model} 
\vspace{-2mm}
\footnotesize{
\begin{tabular*}{\linewidth}{cccc}
\hline\hline\hline
    Replacement Quantity & AUC & HR@5 & NDCG@5\\
\hline
    Replaced one & 0.8735 & 0.5879 & 0.1382\\
    Replaced two & 0.8732 & 0.5866 & 0.1379\\
    Replaced three & 0.8731 & 0.5866 & 0.1377\\
    Replaced five & 0.8731 & 0.5866 & 0.1376\\
    Replaced ten & 0.8730 & 0.5864 & 0.1376\\
\hline\hline\hline
\end{tabular*}
\\\vspace{1mm}\parbox{8.3cm}{Note: We use the DCN model as the base model and use the Amazon (Electro) dataset.}
}
\end{table}

Based on \figref{fig_3}, it is evident that training the noise recognition model with varying number of sequences, corresponding to different quantities of $b_{j}^{D_{(-)}}$ as described in \subsecref{section 3.3.1}, has a different impact on the fitting speed and effectiveness of the noise recognition model. When selecting 0.3 or 0.7 of the data, there is a noticeable decrease in fitting speed and performance. This is due to the use of too few noisy sequences during the training of the noise recognition model, which leads to its inability to accurately recognize noisy sequences. Conversely, if artificially replaced noisy sequences are overly abundant, it can also hinder the noise recognition model's ability to accurately distinguish clean sequences. Furthermore, when opting for a smaller number of sequences, such as using 0.1 of the data as noisy sequences, a situation similar to what was described in \subsecref{section 3.3.1} may occur, namely the noise recognition model struggles to differentiate between noisy and clean sequences. Selecting fewer data points also results in poorer fitting performance.

From \tableref{label_3}, we observe that replacing fewer historical click items in the sequences during noisy recognition model training leads to more accurate output weights for the sequences in the recommender model, resulting in improved performance. However, as the number of replaced items increases, the performance of the recommender model starts to deteriorate. This observation suggests that the original noisy sequences within the historical interaction sequences are mostly sparse. Sequences replacing fewer items exhibit greater similarity to the original noisy sequences, while sequences replacing 3 or more items show significant divergence from the original noisy sequences.

\subsubsection{Sequence Weighting Ablation Experiments (RQ3)}

\renewcommand\arraystretch{1.3}
\begin{table*}[!htb]
\centering
\caption{\label{label_4} Evaluation metrics for different training sequence weighting methods}\vspace{-2mm}
{\footnotesize
\begin{tabular*}{\textwidth}{@{\extracolsep{\fill}}ccccccc}
\hline\hline\hline
    Method & \hspace{-0.4mm}AUC & \hspace{-0.4mm}+RI (AUC) & \hspace{-0.4mm}HR@5 & \hspace{-0.4mm}+RI (HR@5) & \hspace{-0.4mm}NDCG@5 & \hspace{-0.4mm}+RI (NDCG@5) \\
    \hline
    Base (w/o Auxiliary Task or Joint Training) & 0.8729 & - & 0.5865 & - & 0.1376 & - \\
    WAT & 0.8731 & 0.02\% & 0.5862 & -0.05\% & 0.1377 & 0.07\% \\
    DNR & 0.8732 & 0.04\% & 0.5878 & 0.22\% & 0.1381 & 0.36\% \\
    ATJT & 0.8735 & 0.06\% & 0.5879 & 0.24\% & 0.1382 & 0.44\% \\
\hline\hline\hline
\end{tabular*}
\\\vspace{1mm}\parbox{17.5cm}{Note: We use the DCN model as the base model and use the Amazon (Electro) dataset.}
}
\end{table*}

To investigate the impact of different noisy sequence weighting methods on the performance of the recommender model, we compare three weighting training methods: Without Auxiliary Task (WAT), Direct Noise Recognition (DNR) and ATJT. The sequence weights for three methods are obtained through a consistent model structure as described in \subsecref{section 4.1.3}. In the DNR method, the noise recognition model is first separately trained to accurately distinguish artificially replaced noisy sequences from the original clean sequences as the training objective. Then, the training sequences of the recommender model are passed through this noise recognition model to obtain sequence weights, followed by training the recommender model. In the WAT method, the training sequences of the recommender model are directly passed through the targetless model structure to obtain the sequence weights, followed by training the recommender model. 

As shown in \tableref{label_4}, the ATJT method outperforms the other two methods, showing significant improvements in AUC, HR@5 and NDCG@5 metrics. The reason for this is, in comparison to the WAT weighted training method, the ATJT method takes recognition of noise as the auxiliary goal helps the training sequences of the recommender model find suitable training weights. Specifically, it assigns larger weights to clean sequences and smaller weights to noisy sequences, thus mitigates the impact of noisy sequences on the performance of the recommender model. In contrast to the DNR method, the ATJT method not only identifies noisy sequences but also assigns appropriate weights to them. In other words, when training the recommender model with noisy sequences, the goal is not to minimize the weights assigned to them. Rather, the objective is to find the weights that optimizes the performance of the recommender model. It is worth noting that using the ATJT method and the DNR method result in better performance compared with the base model and model trained using the WAT method. This underscores the meaningfulness of weighting sequences through noise recognition auxiliary Task. Additionally, the reason for the inferior performance of the recommender model trained using the WAT method compared with the base model is that the WAT method fails to capture the degree of noise in the sequences. It exhibits confusion in the early stages of training, potentially leading to incorrect weights. This also indicates that augmenting the complexity of the model does not lead to a significant improvement in performance.

\subsubsection{Impact of Noise on Sequence Weights $(w_{i})$}\label{section 4.2.4}

\begin{table}[H]
    \centering 
    \captionof{table}{\label{label_5} Analysis of the weights $w_{i}$ in \equref{eqn_1} generated by the noise recognition model} 
\vspace{-2mm}
\footnotesize{
\begin{tabular*}{\linewidth}{c@{\hspace{0.8cm}}c@{\hspace{0.8cm}}c}
\hline\hline\hline
    Dataset & Sequences & $w_{i}$\\
    \hline
    \multirow{2}{*}{MovieLens20M} 
    & Noisy sequences & $0.4437 \pm 0.0283$\\
    & Clean sequences & $0.5575 \pm 0.0188$\\
    \hline
    \multirow{2}{*}{Amazon (Electro)} 
    & Noisy sequences & $0.4786 \pm 0.0094$\\
    & Clean sequences & $0.5209 \pm 0.0104$\\
    \hline
    \multirow{2}{*}{Yelp} 
    & Noisy sequences & $0.4471 \pm 0.0369$\\
    & Clean sequences & $0.5494 \pm 0.0068$\\
\hline\hline\hline
\end{tabular*}
\\\vspace{1mm}\parbox{8.3cm}{Note: We use the DCN model as the base model. The average and variance of $w_{i}$ for clean and noisy sequences across different datasets.}
}
\end{table}

To demonstrate that noisy sequences have smaller weights compared with clean sequences, we use the DCN model as the base model and conduct analyses on three datasets. As shown in \tableref{label_5}, the weights of the noisy sequences are approximately 19\text{\%} on average smaller compared with the weights of the clean sequences. This aligns with our expectation that assigning smaller weights to noisy sequences can enhance the base model's performance.

\section{Conclusion and Future Work}\label{section 5}
In this study, we proposed a novel self-supervised ATJT method. This method leverages the training outcomes of a noise recognition model to reweight sequences for training the recommender model. Additionally, we conducted joint training of the recommender model and noise recognition model to acquire more appropriate weights, further enhancing the performance of the recommender model. We then evaluated our method on three datasets and six base models, demonstrating its effectiveness. Finally, we validated the impact of different noisy sequences and training methods on recommender model performance through Noise Generation Analysis and Sequence Weighting Ablation experiments.

In the context of future prospects, through adversarial networks, the well-trained noise recognition model can discriminate between artificially replaced noisy sequences, is used as a discriminator to learn a generator that makes it unable to recognize whether the sequences has been artificially replaced with noise. At this point, the generator can create noisier sequences than those replaced by humans, which may be more similar to the noisy sequences in the original sequences. Therefore, using these generated sequences as noisy sequences might yield better results.

\vspace{5mm}

\label{last-page}
\end{multicols}
\end{document}